\documentclass[prl,aps,preprint,superscriptaddress,showpacs]{revtex4}

\usepackage{graphicx}
\begin{document}

\title{Trapping of Cold Excitons with Laser Light}

\author{A.\,T. Hammack}
\author{M. Griswold}
\author{L.\,V. Butov}
 \affiliation{Department of Physics,
University of California at San Diego, La Jolla, CA 92093-0319}

\author{L.\,E. Smallwood}
\author{A.\,L. Ivanov}
\affiliation{Department of Physics and Astronomy, Cardiff
University, Cardiff CF24 3YB, United Kingdom}

\author{A.\,C. Gossard}
\affiliation{Materials Department, University of California at Santa
Barbara, Santa Barbara, California 93106-5050}

\date{\today}

\begin{abstract}
Optical trapping and manipulation of neutral particles has led to a
variety of experiments from stretching DNA-molecules to trapping and
cooling of neutral atoms. An exciting recent outgrowth of the
technique is an experimental implementation of atom Bose-Einstein
condensation. In this paper, we propose and demonstrate laser
induced trapping for a new system---a gas of excitons in quantum
well structures. We report on the trapping of a highly degenerate
Bose gas of excitons in laser induced traps.
\end{abstract}


\maketitle

Lasers enable a precise and non-invasive application of force while
also providing high speed control of the trapping field. This allows
\emph{in-situ} trapping and control for a rich variety of small
neutral particles. Since their origin three decades ago, laser based
traps have been key devices in the advancement of atomic physics and
biophysics, for reviews see
\cite{Chu:1998,Cohen-Tannoudji:1998,Phillips:1998,Ashkin:2000}.

In biology, the applications of optical dipole traps, also known as
optical tweezers, enable direct \emph{in-vivo} manipulation of
viruses, cells, and even individual organelles within the cells. The
use of optical tweezers has also enabled probing of the mechanical
properties of DNA and the forces applied by various molecular motors
found in cells \cite{Ashkin:2000,Lang:2003}.

In atomic physics, the use of the Doppler cooling technique
\cite{Hansch:1975} by sets of counter propagating lasers is employed
to form an ``optical molasses'' containing atoms viscously confined
at microkelvin temperatures. The introduction of optical tweezers to
this molasses enabled the first 3D stable trap for atoms by
capturing them from the surrounding molasses. Following this initial
trapping of atoms, much work was devoted to the creation of larger
volume magneto-optical traps (MOT) to enhance the achievable
densities of trapped atoms
\cite{Chu:1998,Cohen-Tannoudji:1998,Phillips:1998,Ashkin:2000}. It
was specifically these MOTs that led to the first realizations of
Bose-Einstein condensation (BEC) in atoms
\cite{Cornell:1995,Hulet:1995,Ketterle:1995}. Since this initial
realization of BEC, interest has returned again to optical traps,
which can be used to study magnetic effects on BEC, such as Feshbach
resonances, without the complexity added by disrupting the magnetic
field used in MOTs
\cite{Chu:1998,Cohen-Tannoudji:1998,Phillips:1998,Ashkin:2000}. The
possibility of patterning and controlling the potential profile by
laser excitation is also effectively employed in studies of atom BEC
in optical lattices \cite{Greiner:2002}.

In this paper, we propose and demonstrate laser induced trapping for
a new system - a gas of excitons in coupled quantum wells (CQW).
Since the quantum degeneracy temperature scales inversely with the
mass, quantum exciton gases can be achieved at temperatures of about
1 K \cite{Keldysh:1968}, several orders of magnitude higher than
quantum atom gases \cite{Cornell:2002,Ketterle:2002}. Indeed, the
transition from a classical to quantum gas occurs when bosons are
cooled to the point where the thermal de Broglie wavelength
$\lambda_{\rm dB}=\sqrt{2\pi \hbar^2/(mk_BT)}$ is comparable to the
interparticle separation (for instance, BEC takes place when
$n\lambda_{\rm dB}^3=2.612$ in 3D systems) and the transition
temperature for excitons in GaAs/AlGaAs QWs reaches a value of
$T_{\rm dB}=2\pi \hbar^2n_{\rm 2d}/(mgk_B) \approx 3$ K for the
exciton density per spin state $n_{\rm 2d}/g=10^{10}$cm$^{-2}$ (the
exciton spin degeneracy $g=4$ and the exciton mass $m=0.22m_0$ for
GaAs/AlGaAs QWs \cite{Butov:2004}, where $m_0$ is the free electron
mass). Because of their long lifetime and high cooling rate,
indirect excitons in a CQW (Fig. 1a) form a system where a cold and
dense exciton gas can be created, with temperatures well below 1 K
and densities above $10^{10}$ cm$^{-2}$ \cite{Butov:2004}.
Therefore, we chose indirect excitons for development of a method to
trap cold excitons with laser light. This technique opens a pathway
towards high speed control of quantum gases of bosons in
semiconductors---quantum exciton gases.

The possibility of exciton confinement and manipulation in
potential traps attracted considerable interest in earlier
studies. Pioneered by the electron-hole liquid confinement in the
strain-induced traps \cite{Wolfe1975}, exciton confinement has
been implemented in various traps: strain-induced traps
\cite{Trauernicht1983,Kash1988}, traps created by laser-induced
local interdiffusion \cite{Brunner1992}, magnetic traps
\cite{Christianen1998}, and electrostatic traps
\cite{Zimmermann1997,Huber1998,Hammack2005}.

The principle underlying the new method of laser induced exciton
trapping is described below. The CQW geometry~\cite{Butov:2004} is
engineered so that the interaction between excitons is repulsive:
Indirect excitons, formed from electrons and holes that are confined
to different QWs by a potential barrier, behave as dipoles oriented
perpendicular to the plane, and an increasing exciton density causes
an increase of the interaction energy
\cite{Yoshioka,Zhu,Ivanov2002}. The repulsive character of the
interaction is evidenced in experiment as a positive and monotonic
line shift with increasing density \cite{Butov:2004}. Due to the
repulsive interaction, a ring-shaped laser spot should form a
potential trap with the energy minimum at the ring center. Similarly
to all optical traps, an important advantage of the laser induced
exciton trapping is the possibility of controlling the trap
\emph{in-situ} by varying the laser intensity in space and time.
Moreover, the excitons at the trap center are cold since they are
far from the hot laser excitation ring. The long lifetimes of the
indirect excitons allow them to travel to the trap center, due to
their drift and diffusion, before optical recombination. This leads
to accumulation of a cold and dense exciton gas at the trap center.
The implementation of this idea is described below.

In our experiments, the spatial $x$-$y$ photoluminescence (PL)
pattern is acquired by a nitrogen-cooled CCD camera after spectral
selection by an interference filter chosen to match the indirect
exciton energy exclusively. As a result, we are able to remove the
low-energy bulk emission that otherwise dominates the spectrum under
the laser excitation area. This allows the direct visualization of
the indirect exciton PL emission intensity profile in spatial
coordinates (see Figs.~1c-e). In addition, in Fig.~1b we plot the
exciton PL in the \emph{energy-coordinate} plane as measured when a
slit along the diameter of the ring is dispersed by a spectrometer
without spectral selection by an interference filter. Our
investigations determined that a laser excitation ring with a
diameter of 30\,$\mu$m and a ring thickness following a Gaussian
profile of FWHM $= 2 \sigma \simeq 7$\,$\mu$m provided the optimal
conditions for our CQW sample. The ring shaped cw laser excitation
was performed by Nd:YVO$_4$ laser at 532\,nm, or HeNe laser at
633\,nm, or Ti:Sapphire laser tuned to the direct exciton resonance
of 788\,nm. Spatial and spectral features were essentially similar
for all excitation wavelengths investigated. In the experiments with
excitation above the AlGaAs barrier ($\lambda=532$ or 633\,nm),
photoexcited unbalanced charges and the external ring are created,
while in the experiments at nearly resonant excitation
($\lambda=788$ \,nm), no photoexcited unbalanced charges or external
ring are created \cite{Butov:2004}. Comparison of these two
experiments has shown that the charge imbalance and external ring
make no noticeable effect on the exciton trapping. All experimental
data presented here are from a set of 532\,nm excitation ring data
taken with excitation powers $P_{\rm ex}$ in the range
1-1000\,$\mu$W and with gate voltage $V_{\rm g} = 1.4$\,V. The CQW
structure investigated is grown by molecular beam epitaxy and
contains two $8$\,nm GaAs QWs separated by a $4$\,nm
Al$_{0.33}$Ga$_{0.67}$As barrier (details on the CQW structures can
be found in \cite{Butov:2004}). The indirect excitons in the CQW
structure are formed from electrons and holes confined to different
QWs (Fig. 1a). Due to the separation between the electron and hole
layers in the CQW structure, the intrinsic radiative lifetimes of
the optically active indirect excitons exceed that of regular direct
excitons by orders of magnitudes, and are 42 ns at $V_{\rm g} =
1.4$\,V \cite{Butov:2004}.

\begin{figure}[t]
\includegraphics[width=8.5cm]{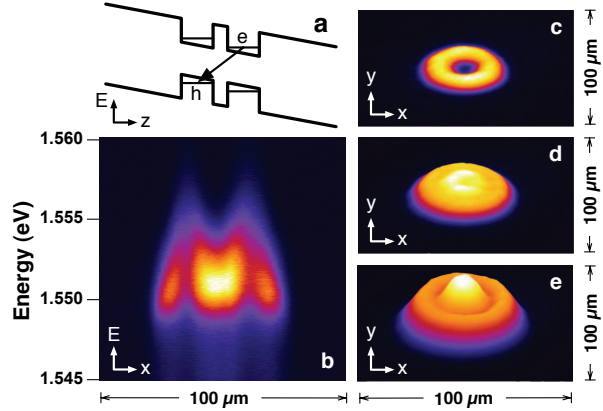}
\caption{Images of laser-induced trapping of excitons. (a) Energy
band diagram of the CQW structure; e, electron; h, hole. (b) Image
of the PL signal in $E$-$x$ coordinates. (c)-(e) Experimental
$x$-$y$ plots of the PL intensity from indirect excitons created by
532\,nm cw laser excitation in a 30\,$\mu$m diameter ring on the CQW
sample. For (c)-(e) the excitation powers are $P_{\rm ex} =$ 10, 35,
100\,$\mu$W and for (b) $P_{\rm ex} =$ 75\,$\mu$W. Sample
temperature $T_{\rm b} = 1.4$\,K.}
\end{figure}

As can be seen in Figs.~1c-e and 2a, for low excitation powers the
PL profile follows the laser excitation ring; however, with
increasing excitation power a spatial PL peak emerges at the center
of the laser excitation ring, indicating the accumulation of a cold
and dense exciton gas. The exciton degeneracy at the trap center
$N_{E=0} = \exp(T_0/T)-1$, where $T_0=2\pi\hbar^2n_{\rm 2d}/(mgk_B)$
\cite{Ivanov1999}, can be estimated from the exciton density and
temperature. The exciton density $n_{\rm 2d} = \varepsilon \delta E
/(4\pi e^2 d)$ is measured directly by the exciton energy shift
$\delta E$, where $d = 12$\,nm is the separation between the
electron and hole layers for our samples and $\varepsilon$ is the
background dielectric constant \cite{Yoshioka,Zhu,Ivanov2002}
($n_{\rm 2d} = 10^{10}$\,cm$^{-2}$ for $\delta E \simeq 1.6$\,meV).
The exciton temperature at the trap center is essentially equal to
the lattice temperature due to absence of heating sources at the
trap center. The estimate shows that for the excitation $P_{\rm
ex}=1000 \mu$W and temperature $T_{\rm b} = 1.4$\,K, see the
experimental data in Fig. 2, the exciton degeneracy at the trap
center is $N_{E=0} \simeq 8$. The theoretical modeling presented
below confirms this estimate.

\begin{figure}[t]
\includegraphics[width=8.5cm]{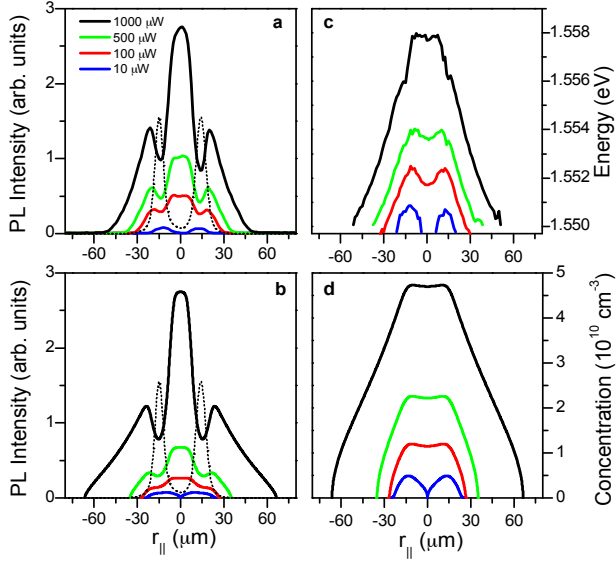}
\caption{Spatial profiles of the PL intensity and energy for the
excitons in the laser-induced trap. (a) Measured PL intensity, (b)
calculated PL intensity, (c) measured energy position of the PL
line, and (d) calculated exciton concentration against radius
$r_{\|}$ for four optical excitation powers $P_{\rm ex}$. The
vertical axes of (c) and (d) cover the same range due to the
relation $\delta E = 4\pi e^2 n_{\rm 2d} d /\varepsilon$, where $d =
12$\,nm for our sample. The ring shaped profile of the laser
excitation is shown by the thin dotted lines in (a) and (b). Sample
temperature $T_{\rm b} = 1.4$\,K.}
\end{figure}

Note that the exciton trapping by laser light is based on a
different physical principle compared to the atom trapping by laser
light. However, the two techniques lead to conceptually similar
optical trapping of quantum gases---of excitons or atoms,
respectively.

A parabolic energy trap is apparent in the interior of the
excitation ring (Figs.~1b and 2c). The decrease in the indirect
exciton PL at the location of the excitation ring (Fig. 2a) is
because the high-energy photogenerated excitons heat the exciton
gas; this heating reduces the fraction of optically active excitons,
which have low energies, $E \le E_0=E_g^2\varepsilon/(2mc^2)$ where
$E_g$ is the energy gap and $c$ is the speed of light
\cite{Feldman87}. As they drift and diffuse away from the excitation
area, the excitons thermalize to the lattice temperature $T_{\rm b}$
and become optically active, leading to the moderately enhanced PL
intensity directly external to the excitation ring. The strong
enhancement of the PL at the excitation ring center, Fig. 2a, is due
to (1) the excitons' thermalization to the lattice temperature and
(2) the accumulation of large numbers of excitons driven by dipole
repulsion away from the higher density region towards the ring
center.

The numerical simulations, based on a microscopic theoretical model,
match the experimental results excellently (Fig.~2b and d). Our
approach to the transport, relaxation and PL dynamics of indirect
excitons is formulated in terms of three coupled nonlinear
equations: A quantum diffusion equation for the exciton density
$n_{\rm 2d}$, a thermalization equation for the exciton temperature
$T$, and an equation for the exciton optical lifetime $\tau_{\rm
opt}$ \cite{Ivanov2002}. Calculation details will be provided
elsewhere. A particular feature of the trap is that it is formed by
the indirect excitons themselves: The trap potential is given in the
mean-field approximation by $U_{\rm trap} = \delta E = u_0 n_{\rm
2d} = 4\pi e^2 n_{\rm 2d} d /\varepsilon$, where $u_0$ is a positive
scattering amplitude. Note that the trap confining potential is
determined by the radial exciton density distribution and is
essentially independent of other characteristics of indirect
excitons such as their temperature, etc.. Quantum-statistical
corrections \cite{Ivanov2002}, which enhance the nonlinear effects,
are included in the simulations.

\begin{figure}[t]
\includegraphics[width=4.4cm]{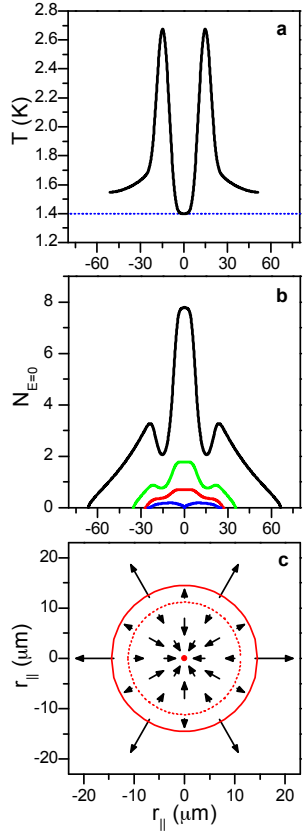}
\caption{The calculated parameters of excitons in the laser-induced
traps. (a) The radial dependence of the exciton temperature $T$ for
optical excitation power of 1000\,$\mu$W (solid black) and the
lattice temperature $T_{\rm b} =$ 1.4\,K (dotted blue). (b) The
occupation number $N_{E=0}$ for our samples against radius $r_{\|}$
for the same four optical excitation powers as in Fig.~2. (c) Vector
plot showing the calculated exciton velocities $v$ for excitation
power of 500\,$\mu$W. The length of the outermost vectors is $1.3
\times 10^6$\,cm/s. The dotted red line shows the maximum calculated
exciton concentration and the solid red line shows the maximum of
incident laser power used in calculation.}
\end{figure}

The increase of the exciton gas temperature due to heating by
photogenerated excitons at the excitation ring is evident in
Fig.~3a. A minor heating due to the exciton potential energy
gradient can also be seen outside the excitation ring. The exciton
transport towards the trap center due to drift and diffusion is
illustrated in Fig.~3c. Finally, the theoretical calculations
confirm that in our experiments, which deal with the cryostat
temperature $T_{\rm b} = 1.4$\,K, high nonclassical occupation
numbers, $N_{E=0} \simeq 8$, build up at the trap center (Fig.~3b).

In summary, in this paper we propose and demonstrate a method to
trap cold exciton gases with laser light. The laser induced exciton
trapping makes it possible to control the trap \emph{in-situ} by
varying the laser intensity in space and time. The excitons at the
trap center are cold since they are far from the hot laser
excitation ring. We report on the trapping of a cold gas of excitons
in a laser induced trap and on the formation of a highly degenerate
Bose gas of excitons in the trap.

We thank L.V. Keldysh, L.J. Sham, and D.E. Smith for valuable discussions.
This work was supported by ARO (Grant No. W911NF-05-1-0527).

%\bibliography{references}

\end{document}